%% file: paper-eurosec.tex
\documentclass[sigconf,authorversion=true]{acmart}

\usepackage{booktabs} % For formal tables

\usepackage[utf8]{inputenc}
\usepackage{multirow}
\usepackage{microtype}
\usepackage{todonotes}

\usepackage{subcaption}

\usepackage[toc,acronym,nowarn]{glossaries}
\input{gloss-acr}
\makeglossaries
\glsdisablehyper % disable hyperlinks to nowhere...
\usepackage{xspace}
\newcommand*{\eg}{e.g.,\@\xspace}
\newcommand*{\ie}{i.e.,\@\xspace}

\newcommand{\name}{SEVered\xspace}
\newcommand{\Title}{\name: Subverting AMD's Virtual Machine Encryption}

\usepackage{graphicx}
\graphicspath{{graphics/}}
\DeclareGraphicsExtensions{.pdf,.jpeg,.png}

\usepackage{hyperref}
\hypersetup{
  colorlinks=true, linkcolor=black, citecolor=black, urlcolor=black,
  pdftitle={\Title}
  pdfauthor={Mathias Morbitzer, Manuel Huber, Julian Horsch, Sascha Wessel}
}

\usepackage[all]{hypcap}
\usepackage{fixltx2e}
\usepackage{enumitem} % fixes overfull hboxes in description envs
\usepackage{nicefrac}

\copyrightyear{2018}
\acmYear{2018}
\setcopyright{rightsretained}
\acmConference[EuroSec'18]{11th European Workshop on Systems Security}{April 23--26, 2018}{Porto, Portugal}
\acmBooktitle{EuroSec'18: 11th European Workshop on Systems Security, April 23--26, 2018, Porto, Portugal}
\acmDOI{10.1145/3193111.3193112}
\acmISBN{978-1-4503-5652-7/18/04}

\begin{document}

%\acmArticle{4}
%\acmPrice{15.00}

% These commands are optional
%\acmBooktitle{Transactions of the ACM Woodstock conference}
%\editor{Jennifer B. Sartor}
%\editor{Theo D'Hondt}
%\editor{Wolfgang De Meuter}

\title{\Title}

\author{Mathias Morbitzer, Manuel Huber, Julian Horsch and Sascha Wessel}
\affiliation{
  \institution{Fraunhofer AISEC}
  \city{Garching near Munich}
  \state{Germany}
}
\email{{firstname.lastname}@aisec.fraunhofer.de}

\begin{abstract}
\input{abstract}
\end{abstract}

%
% TODO



%\keywords{mobile device security, security architecture, data confidentiality, operating system security}

\maketitle

\input{content}

\bibliographystyle{ACM-Reference-Format}
\bibliography{biblio}

\end{document}

%% file: gloss-acr.tex
\newacronym{OS}{OS}{Operating System}
\newacronym{SE}{SE}{Secure Element}
\newacronym{NFC}{NFC}{Near Field Communication}
\newacronym{BLE}{BLE}{Bluetooth Low Energy}
\newacronym{RIL}{RIL}{Radio Interface Layer}
\newacronym{SELinux}{SELinux}{Security-Enhanced Linux}
\newacronym{LSM}{LSM}{Linux Security Modules}
\newacronym{TCB}{TCB}{Trusted Computing Base}
\newacronym{MAC}{MAC}{Mandatory Access Control}
\newacronym{IPC}{IPC}{Inter-Process Communication}
\newacronym{ICC}{ICC}{Inter-Container Communication}
\newacronym{cgroups}{cgroups}{control groups}
\newacronym{CA}{CA}{Certificate Authority}
\newacronym{PKI}{PKI}{Public Key Infrastructure}
\newacronym{MITM}{MITM}{Man-In-The-Middle}
\newacronym{BYOD}{BYOD}{Bring-Your-Own-Device}
\newacronym{CM}{CM}{Container Management}
\newacronym{SM}{SM}{Security Management}
\newacronym{HAL}{HAL}{Hardware Abstraction Layer}
\newacronym{TLS}{TLS}{Transport Layer Security}
\newacronym{C2C}{C2C}{Container To Container}
\newacronym{Protobuf}{Protobuf}{Protocol Buffers}
\newacronym{SDO}{SDO}{Sensitive Data Object}
\newacronym{VMA}{VMA}{Virtual Memory Area}
\newacronym{PGD}{PGD}{Page Global Directory}
\newacronym{PUD}{PUD}{Page Upper Directory}
\newacronym{PMD}{PMD}{Page Middle Directory}
\newacronym{PTE}{PTE}{Page Table Entry}
\newacronym{COW}{COW}{Copy-On-Write}
\newacronym{IV}{IV}{Initialization Vector}
\newacronym{ESSIV}{ESSIV}{Encrypted Salt-Sector Initialization Vector}
\newacronym{KSM}{KSM}{Kernel Samepage Merging}
\newacronym{JIT}{JIT}{Just-In-Time}
\newacronym{DMA}{DMA}{Direct Memory Access}
\newacronym{FDE}{FDE}{Full Disk Encryption}
\newacronym{AS}{AS}{Address Space}
\newacronym{GCM}{GCM}{Google Cloud Messaging}
\newacronym{TPM}{TPM}{Trusted Platform Module}
\newacronym{JTAG}{JTAG}{Joint Test Action Group}
\newacronym{LUKS}{LUKS}{Linux Unified Key Setup}
\newacronym{VPN}{VPN}{Virtual Private Network}
\newacronym{PBKDF2}{PBKDF2}{Password-Based Key Derivation Function 2}
\newacronym{KVM}{KVM}{Kernel-based Virtual Machine}
\newacronym{VM}{VM}{Virtual Machine}
\newacronym{HV}{HV}{Hypervisor}
\newacronym{SEV}{SEV}{Secure Encrypted Virtualization}
\newacronym{SME}{SME}{Secure Memory Encryption}
\newacronym{SP}{SP}{Secure Processor}
\newacronym{GVA}{GVA}{Guest Virtual Address}
\newacronym{GPA}{GPA}{Guest Physical Address}
\newacronym{HPA}{HPA}{Host Physical Address}
\newacronym{GPT}{GPT}{Guest Page Table}
\newacronym{HPT}{HPT}{Host Page Table}
\newacronym{TLB}{TLB}{Translation Lookaside Buffer}
\newacronym{PoC}{PoC}{Proof of Concept}
\newacronym{ORAM}{ORAM}{Oblivious RAM}
\newacronym{SEV-ES}{SEV-ES}{SEV Encrypted State}
\newacronym{VMCB}{VMCB}{Virtual Machine Control Block}
\newacronym{SLAT}{SLAT}{Second Level Address Translation}

%% file: abstract.tex
AMD SEV is a hardware feature designed for the secure encryption of virtual machines.
SEV aims to protect virtual machine memory not only from other malicious guests and physical attackers, but also from a possibly malicious hypervisor.
This relieves cloud and virtual server customers
from fully trusting their server providers and the hypervisors they are
using.
We present the design and implementation of \name, an attack from a malicious
hypervisor capable of extracting the full contents of main memory in plaintext from SEV-encrypted virtual machines.
\name neither requires physical access nor colluding virtual
machines, but only relies on a remote communication service, such as a web server,
running in the targeted virtual machine.
We verify the effectiveness of \name on a recent AMD SEV-enabled server platform running different services, such as web or SSH servers, in encrypted virtual machines.
With these examples, we demonstrate that \name reliably and efficiently extracts all memory
contents even in scenarios where the targeted
virtual machine is under high load.

%% file: content.tex
\section{Introduction}
\label{sec:introduction}

As a common practice, cloud and virtual server customers conveniently run their services in \glspl{VM} remotely operated on the platforms of their server providers.
The privileged \glspl{HV} on these platforms ensure the logical separation of multiple \glspl{VM} operating on the same hardware.
Attackers have demonstrated that they are capable of circumventing this
protection, achieving access to the memory of the \glspl{VM}, \eg with memory
attacks via Coldboot \citep{Halderman08lestwe} or \gls{DMA}
\citep{boileau2006hit, pcie, becher2005firewire}, or even gaining complete control of the \gls{HV} \cite{Microsoft2017Vuln, VMWare2017Vuln, Xen2017Vuln}.
However, the server provider running the \gls{HV} poses the most obvious danger to the \gls{VM}'s integrity and data confidentiality.
Customers must rely on the trustworthiness of the providers and their
\glspl{HV}, since they
can easily access all of the \gls{VM}s' memory and identify sensitive data such as keys, passwords, or classified information.

To reduce the attack surface of virtualized systems towards malicious server providers, AMD introduced \gls{SEV} \cite{AMD2017API}.
\gls{SEV} is capable of transparently encrypting individual \glspl{VM} using a \gls{SP}.
The technology especially targets server systems and enables \glspl{VM} to
request encryption and receive proof about the encryption from the \gls{SP}.
The memory of each protected \gls{VM} is encrypted within the \gls{SP} based on an individual ephemeral key never leaving the \gls{SP}.
The implementation in hardware not only makes the systems resistant against memory attacks, but also prevents \glspl{HV} from accessing sensitive \gls{VM} data.

With \name, we demonstrate that it is nevertheless possible for a malicious \gls{HV} to extract all memory of an \gls{SEV}-encrypted \gls{VM} in plaintext.
We base \name on the observation that the page-wise encryption of main memory lacks integrity protection \cite{kaplan2016amd, AMD2017API, hetzelt2017security}.
While the \gls{VM}'s \gls{GVA} to \gls{GPA} translation is controlled by the
\gls{VM} itself and opaque to the \gls{HV},
the \gls{HV} remains responsible for the \gls{SLAT}, meaning that it maintains
the \gls{VM}'s \gls{GPA} to \gls{HPA} mapping in main memory.
This enables us to change the memory layout of the \gls{VM} in the \gls{HV}.
We use this capability to trick a service in the \gls{VM}, such as a web server, into returning arbitrary pages of the \gls{VM} in plaintext upon the request of a resource from outside.
We first identify the encrypted pages in memory corresponding to the resource,
which the service returns as a response to a specific request.
By repeatedly sending requests for the same resource to the service while
re-mapping the identified memory pages, we extract all the \gls{VM}'s memory
in plaintext.
\name neither requires detailed knowledge of the target \gls{VM} or
service, nor a malicious process colluding from inside the \gls{VM}. Our attack
is also resistant to noise, \ie concurrent activity in the target \gls{VM},
and dynamically adapts to different noise levels.

\section{Attack Method}
\label{sec:design}

\begin{figure}[t]
  \centering
  \includegraphics[width=0.9\linewidth]{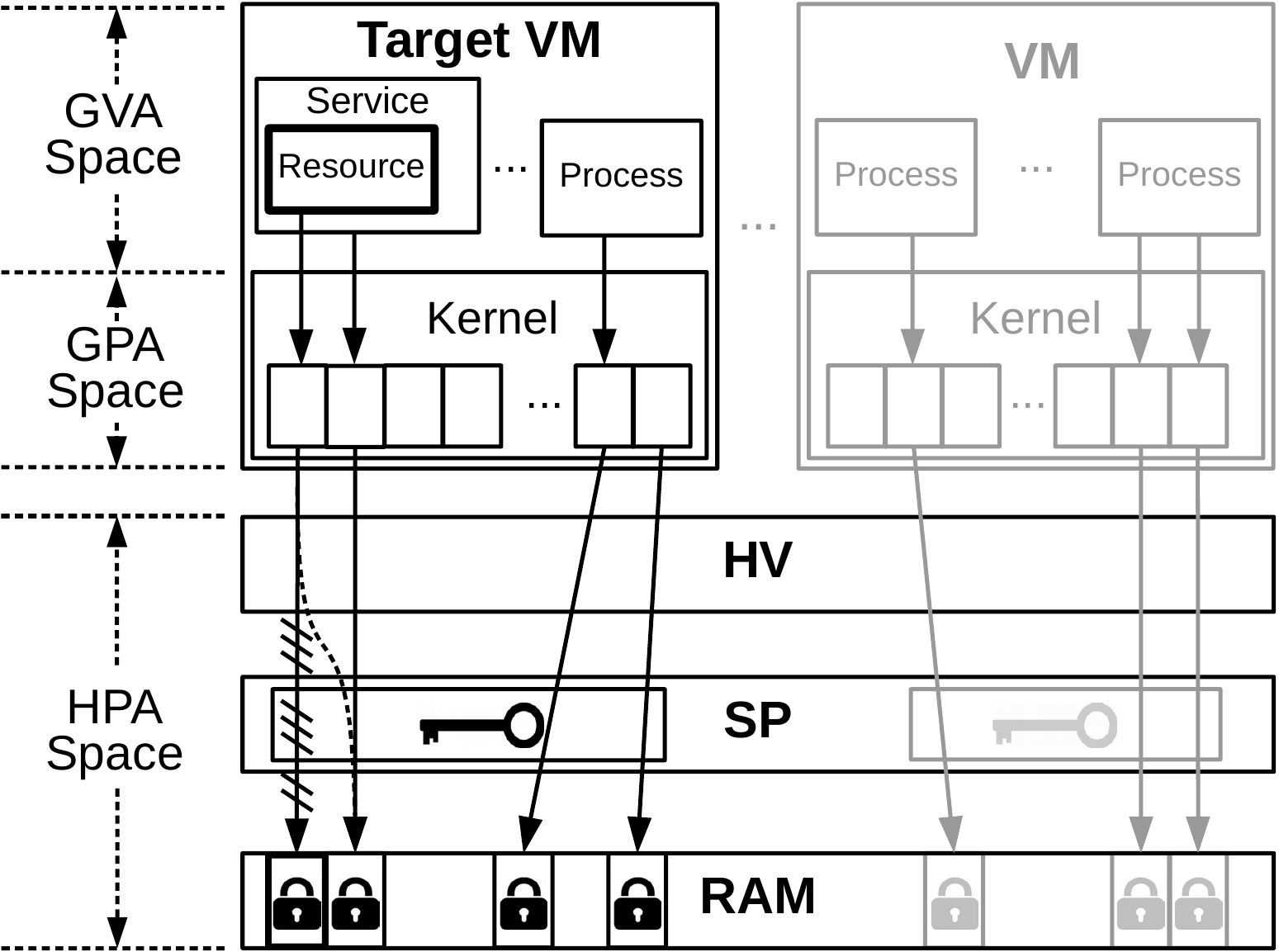}
  \caption{Overview on our memory extraction concept on a memory encryption platform with different \glspl{VM}.}
  \label{fig:architecture}
\end{figure}

Our target is an AMD SEV-enabled platform which runs an
attacker-controlled \gls{HV} and one or more \glspl{VM} as shown in Figure~\ref{fig:architecture}. 
Our target \gls{VM}'s memory is fully encrypted by \gls{SEV}. 
While being able to target multiple \glspl{VM} at the same time, we describe
our attack for a single \gls{VM}.
Inside the target \gls{VM}, we assume presence of the following
\textbf{components}:
\begin{description}
	\item[Service.] A process running inside the \gls{VM} 
		offering a \emph{resource} via a publicly accessible remote connection.
		Common examples are HTTP, SSH, FTP or mail servers.
	\item[Resource.] Data in \gls{VM} memory that is remotely readable through a
		\emph{service}. A resource can spread over one or more memory
		pages. A representative example in the context of a web
		server is a HTML page or a file offered for download.
		The suitability of a resource for \name depends on its
		\emph{size} and \emph{stickiness}, as discussed in Sections~\ref{sec:design:resource_size} and
		\ref{sec:design:resource_stickiness}.
		We consider a resource to be \emph{sticky} if it is probable that the resource
		remains present in guest-physical memory and is not relocated or evicted during our attack.

\end{description}
After identifying an appropriate service and resource in the target \gls{VM},
the actual \gls{VM} memory extraction is executed.  The following two
\textbf{phases} characterize our method:
\begin{enumerate}
\item \textbf{Resource Identification.} In this phase, we identify memory pages of the
	chosen resource in physical memory. As described
	before, the address translation in the \gls{VM} is opaque
	to the \gls{HV} due to \gls{SEV}'s memory encryption. Therefore, several techniques have to be
	combined to reliably identify the \glspl{GPA} belonging to the
	resource, as described in Section~\ref{sec:design:identifying}.
\item \textbf{Data Extraction.} In this phase, we extract
	plaintext from the encrypted \gls{VM} by repeating
	requests while switching the mapping of the identified resource \glspl{GPA} to
	\glspl{HPA} in the \gls{HV} as shown on the bottom of \autoref{fig:architecture}.
	After completion, we restore the original state of the \gls{VM} by
	mapping the resource pages to their original \glspl{HPA}.
	This phase is described in Section~\ref{sec:design:extraction}.
\end{enumerate}

\subsection{Resource Identification}
\label{sec:design:identifying}

Despite the opaqueness of the \gls{VM}'s guest memory mappings due to \gls{SEV}'s encryption,
the \gls{HV} still controls the mapping from \glspl{GPA} to actual physical
pages, \ie \glspl{HPA}. We use this capability to establish a basic technique
we call \emph{page tracking} to gain information about the \gls{VM}'s
memory layout. During page tracking our \gls{HV} registers each
access to a \gls{GPA} and the corresponding physical page by the targeted \gls{VM}. 
We realize the page tracking 
by invalidating the \glspl{PTE} of the \gls{VM}, \ie by removing their
\emph{present} flags. As soon as a page is accessed, it triggers a page fault which
we record before setting the present flag again. Hence, the page tracking
triggers exactly \emph{once} for each accessed page before tracking is restarted.

The goal of the resource identification phase is to identify the set of pages that store the service's
response, \ie the target resource. We define the unknown set of resource pages as
follows:
\[ R = \{p : \textrm{Page } p \textrm{ contains (part of) the target resource}\} \]

The only information available to identify the target resource
are the \glspl{GPA} accessed by the \gls{VM} as result of the described tracking process
by the \gls{HV}. When simply tracking a request to the target resource, the result
contains a lot of pages that do not belong to the resource. With
additional noise, \eg concurrent activity caused by other clients accessing the \gls{VM}, the
tracking also records those accesses, making the result even fuzzier.
To reliably identify our target resource in \glspl{VM} with varying noise
levels, we propose an iterative approach. The higher the noise, the more
iterations can be conducted to converge to an approximation of $R$. 
We repeat the following steps $n \in
\mathbb{N}$ times. The current iteration of our identification process is $i$
with $1 \leq i \leq n$.
We start each iteration $i$ by requesting the resource via the target service and record all
pages the \gls{VM} accesses while fulfilling the request:
\[ R_i = \{p : \textrm{Page } p \textrm{ accessed during request } i \textrm{ for target
			resource}\} \]
Since we request the resource ourselves, this gives us a sample set which is
\emph{guaranteed} to contain the resource pages:
\[ R \subseteq R_i \] 
$R_i$ is typically large ($|R| \ll |R_i|$) since it
includes not only the resource itself but also other pages of the
service as well as memory pages of other processes and parts of the kernel.
All concurrent activity during the recording must be considered as noise, as
it directly increases $|R_i|$. We call this type of noise \emph{R-noise}.
Based on the observation that $R \subseteq R_i$, we can refine this set by
intersecting all $R_i$ recorded up to this point for which also holds $R \subseteq R^i$:  
\[ R^i = R^{i-1} \cap R_i \qquad R^0 = R_1 \] 
$R^i$ is updated on each iteration and contains only pages that are accessed
for each access to the resource. After an appropriate number of iterations it
should therefore only contain pages directly required for fulfilling the
request to the target resource, filtering unrelated pages, such as pages from
other processes.

Next, we want to sample a set of page accesses which is \emph{similar} to $R_i$
but without accessing our target resource.
Hence, we continue our iteration $i$ by requesting an arbitrary \emph{other resource},
for example, a different web page, from the same service. 
Again, we track and record all pages the \gls{VM} accesses during the request:
\[ X_i = \{p : \textrm{Page } p \textrm{ accessed during request } i \textrm{ for other resource}\} \]
$X_i$ only contains target resource pages if another client accesses the resource
 while we record.
Hence, pages that are part of $X_i$ are \emph{unlikely} to contain $p \in R$.
Based on this, we define a set of \emph{likely candidates} $C_i$ for each iteration by subtracting
those pages from $R^i$:
\[ C_i = R^i \setminus X_i \]
This step filters all pages that are part of the service but not the resource itself. 
Because of the subtraction, while recording $X_i$ we must only consider
accesses to the target resource as noise. We call this \emph{X-noise}.
We define a multiset $C^i$, which provides the information
\emph{how often} a page was identified as candidate, to gather all candidates from all iterations.
We denote the multiplicity of an element $p$ in
a multiset $A$ as $A(p)$. Based on this, we specify the union of a multiset as
sum of multiplicities, \ie $(A \uplus B)(p) = A(p) + B(p)$, and the
intersection as multiplication of multiplicities, \ie $(A \cap B)(p) = A(p)
\cdot B(p)$. With this we define $C^i$ as:
\[ C^i = (C^{i-1} \uplus C_i) \cap R^i \qquad C^0 = \emptyset \]
For candidates in $C_i$ which are already present in $C^{i-1}$, the multiplicity
increases in $C^i$.
The intersection with $R^i$ ensures that candidates  from
a previous iteration ($p \in C^{i-1}$) that can be excluded with the knowledge gained in the current iteration ($p
\notin R^i$) are completely removed from $C^i$.
We calculate the probability that a candidate page $p \in C^i$ is part of our target resource
after iteration $i$ based on how often it was a candidate:
\[ \textrm{P}_i[p \in R] = \frac{C^i(p)}{|C^i|} \]
Note that if $|R| > 1$, the probability is distributed between all $p \in R$.
Therefore, the probability is only interpreted in relation to the probability
of other pages $p \in C^i$.

Finally, after $n$ iterations, we calculate the probability $\textrm{P}_n[p \in R]$ for each $p
\in C^n$ and build a list of candidate pages sorted by probability. 
By choosing $n$ appropriately, our model is able to remove noise during both sampling
phases (R- and X-noise), as shown in our evaluation in \autoref{sec:evaluation}. 
With the resulting list, we start the extraction phase described in the next section.

\paragraph{Example}
In order to clarify the resource identification mechanism, consider the
following simplified example. 
During our first iteration ($i=1$) we record $R_1$ and $X_1$ and calculate:
\begin{flalign*}
 	\qquad\qquad R_1 &= \{4,8,15,16,23,42\} &\\
	R^1 &= R^0 \cap R_1 = R_1 \cap R_1 = \{4,8,15,16,23,42\} &\\
	X_1 &= \{3,8,12,15,16,23,27\} \\
	C_1 &= R^1 \setminus X_1 = \{4,42\} \\
	C^1 &= (C^0 \uplus C_1) \cap R^1 = (\emptyset \uplus \{4,42\}) \cap R^1 = \{4,42\} 
\end{flalign*}
During our second iteration ($i=2$) we record and calculate:
\begin{flalign*}
	\qquad\qquad R_2 &= \{6,8,15,16,23,42\} &\\
	R^2 &= R^1 \cap R_2 = \{8,15,16,23,42\} \\
	X_2 &= \{2,8,12,13,15,23\} \\
	C_2 &= R^2 \setminus X_2 = \{16,42\}\\
	C^2 &= (C^1 \uplus C_2) \cap R^2 = (\{4,42\} \uplus \{16,42\}) \cap R^2\\
	&= \{4,16,42,42\} \cap \{8,15,16,23,42\} = \{16,42,42\}
\end{flalign*}
Assuming $n=2$, we now finished all iterations and calculate $\textrm{P}_2[p \in R]$ for the pages $p \in
C^2$:
\[ \textrm{P}_2[42 \in R] = \frac{C^2(42)}{|C^2|} = \frac{2}{3} \qquad \textrm{P}_2[16 \in R] = \frac{C^2(16)}{|C^2|} = \frac{1}{3} \]
Hence, we start our extraction phase with the page list $[42,16]$ ordered by
probability.

\subsection{Data Extraction}
\label{sec:design:extraction}
\label{sec:design:remapping}
\label{sec:design:reading}

The resource identification resulted in a list of \glspl{GPA} for
the target \gls{VM} which most probably contain the target resource.
The data extraction phase uses this list and the ability of the
\gls{HV} to switch mappings from \glspl{GPA} to \glspl{HPA} to extract
arbitrary decrypted memory from our target \gls{VM}. 

First, we determine the number of pages $r$ that are at least necessary to store the target
resource. The sizes of the target resource $S_r$ and a single page $S_p$ are
known, so that we can determine $r$ as \nicefrac{$S_r$}{$S_p$}.
We then take the first $r$ pages of the probability list and repeat the
following two steps:
\begin{enumerate}
	\item \textbf{Page Remapping.} We modify the \gls{HPT} entries of
		the $r$ pages so that their \glspl{GPA} point to the memory
		pages we want to extract as depicted in Figure~\ref{fig:architecture}.
		After the modification we ensure that the corresponding
		\gls{TLB} entries are flushed for the changes to take effect
		immediately.
	\item \textbf{Data Request.} We request the target resource from our service. % as normal client. 
	       	Since the underlying pages for the resource were
		remapped, the service unintentionally responds with data from
		the pages we chose to extract. 
\end{enumerate}
We repeat both steps remapping the resource \glspl{GPA} to all memory regions
of interest to extract them in plaintext.

\begin{figure*}[t]
  \centering
  \includegraphics[width=\textwidth]{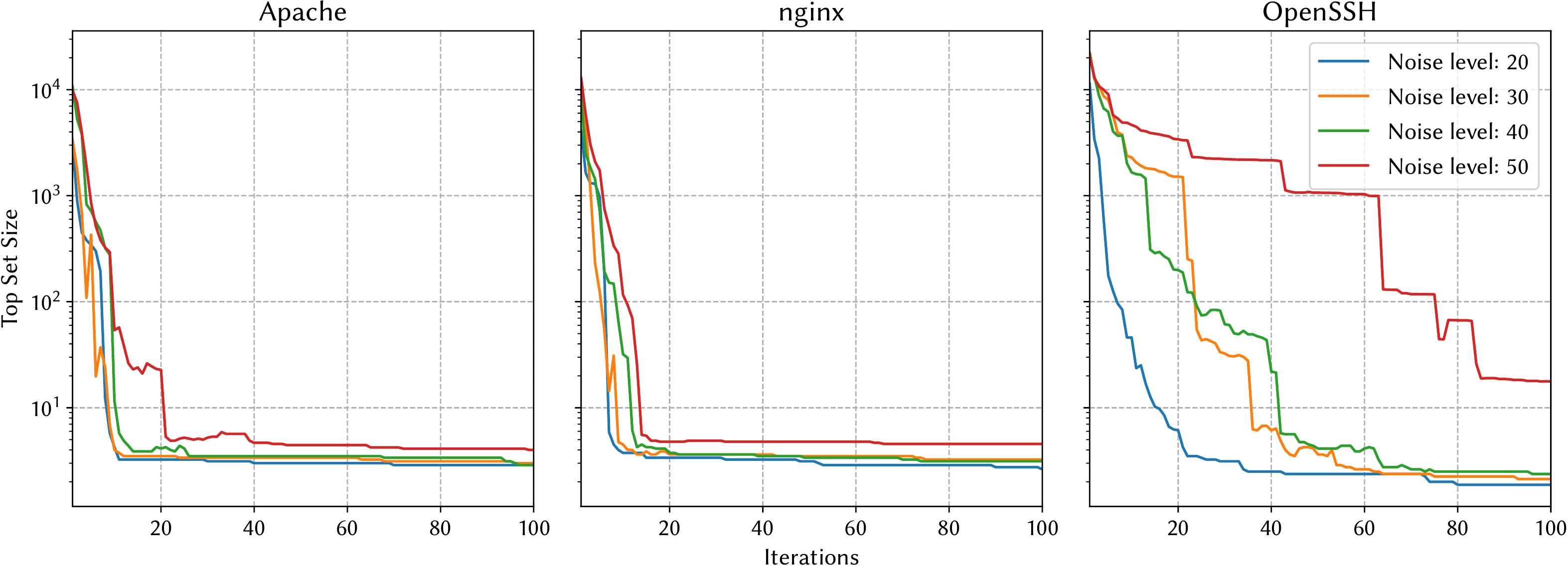}
  \caption{Measurements of top set size $|T|$ for increasing number of iterations and different noise levels.}
  \label{fig:eval_accesses}
\end{figure*}

If we receive the original resource for one or more of the $r$
pages replaced in the first step, those pages do not belong to the target
resource. In this case we continue the extraction with the next 
pages in the probability list or repeat the identification phase.

Concurrent accesses during the extraction phase can have different consequences.
If the target resource itself is
accessed by another client, the access returns wrong data to the client. If a different
resource that \emph{shares} a page with our target resource is accessed, the
access returns wrong data and/or introduces malfunction in the target
\gls{VM}. All other accesses are unaffected. To exclude malfunctions, a target
resource should be chosen which covers one or more complete pages, as
discussed in the following sections.

\subsection{Resource Size}
\label{sec:design:resource_size}
The size of the resource determines whether all memory can be accessed and how many
requests are required during the extraction phase. To access all
areas of memory, the targeted resource must cover at least one entire page.
The resource does \emph{not} have to be page-aligned and can,
for example, cover the second half of one page and the first half of another
page. However, as explained before, such resources should be avoided
because of possible malfunctions when concurrently accessed inside the \gls{VM}.

Since all pages of the target resource can be remapped and used for extraction
in every iteration of the data extraction phase, the larger the resource, the less
iterations are required for extracting
a certain amount of memory.
In our experiments (see Section~\ref{sec:evaluation}) we found that
resources representing the content of a file in memory are especially
convenient.
They have the advantage that a sufficiently large file located in memory
always covers at
least one memory page and that no other data is located on the same page.
This also guarantees that the page is always read starting without offset and
no data is omitted at the beginning.
However, the service only returns as many bytes as the actual size of the file.

\subsection{Resource Stickiness}
\label{sec:design:resource_stickiness}

There are several factors that influence the degree of a resource's stickiness, some of
which we discuss in the following:
\begin{description}
	\item[Resource type.]
	Resource memory pages can have different attributes.
	For example, they can be read-only or also writable. One important
	distinction is, if they are file-backed, \ie are caching a part of a
	file in memory. When the system is low on memory, normal pages are
	swapped and file-backed pages are simply removed from memory. Depending
	on the configuration of the target system, \eg via the
	\texttt{swappiness} file in Linux, file-backed pages are prioritized over
	non-file-backed pages when choosing a page for eviction. Optimally, a resource
	should be chosen based on its type matching the target \gls{VM}'s
	configuration for maximum stickiness.
	\item[Process priority.]
		When the target \gls{VM} completely runs out of memory
		(including swap), it typically starts to kill processes.
		Optimally, a target service should be chosen that is unlikely
		to be killed when running out of memory.
	\item[VM memory pressure.]
		As memory pressure in the \gls{VM} decreases, the stickiness of
		all resources offered by all services in the increases and
		vice versa.
\end{description}

For a target VM running a typical Linux-based OS, we found in our evaluation that
file resources cached in the Linux page cache provide high stickiness.

\section{Evaluation}
\label{sec:evaluation}
\label{sec:impl}

We implemented our prototype on a system powered by an AMD EPYC 7251 processor with SEV fully enabled. 
Our system runs Debian Linux with the SEV-enabled kernel in version \texttt{4.13.0-rc1} and
QEMU \texttt{2.9.50} as provided by AMD~\cite{AMDSEV}.
As malicious \gls{HV}, we used \gls{KVM} and modified it to execute our
attack. To realize our tracking mechanism, we extended the \gls{KVM} infrastructure for guest write
access tracking \cite{guangrong2016patch} to track all kinds of accesses.
We furthermore extended \gls{KVM} with functionality to alter memory mappings for
the extraction phase.
Both features can be controlled by the attacker in the host Linux running the
target \gls{VM}.
In the following, we evaluate our prototype based on services commonly found in target \glspl{VM}.
We chose two web servers, Apache \texttt{2.4.25-3} and nginx \texttt{1.10.3-1}, as well as an SSH server, OpenSSH \texttt{1:7.4p1-10}.
As target resource to be served by these services, we used a file of size 4 KB filling exactly one page in
memory. We evaluate both phases of our attack separately.

\subsection{Resource Identification}

For our evaluation, the target page $tp$ of our page-sized target resource is
known. Based on this, we define a \emph{top set} which we use to measure the performance of our identification
mechanism:
\[ T_i = \{p \in R^i: \textrm{P}_i[p \in R] \geq \textrm{P}_i[tp \in R]\}\]
The top set $T_i$ contains all pages that our identification algorithm considers at least as likely to be the target page as the actual target page $tp$ after $i$ iterations.
The smaller the set $T$, the better the identification.
To ensure that the identification works in real world environments, we
introduced four levels of noise into our target \gls{VM} during our tests. 
A noise level of 20 refers to an environment where on the average 20 random 
accesses per second are made to our three services by arbitrary peers.
Our noise model therefore generates both X-noise and R-noise (see \autoref{sec:design:identifying}).
For each noise level and service, we conducted eight test runs with 100 iterations.

\autoref{fig:eval_accesses} shows the average top set size $|T|$ for each service, noise level, and iteration.
We use a logarithmic scale for the Y-axis, as the top set size quickly decreases after a few iterations.
For Apache and nginx, the top set size quickly converges to about 3 to 4 candidates in average for all noise levels.
The same holds for OpenSSH, where the top set size converges to about 2 to 3 candidates in average except for noise level 50 where 100 iterations were not sufficient to identify a small top set.
The results show that the \name identification mechanism is able to
handle noisy environments by dynamically increasing the number of iterations.
\autoref{tab:time_finished} summarizes the number of iterations and absolute
time required for every service and noise level until the top set converges to
less than 5 candidates.
Even in the highest noise level, the top set converges in less than 23 seconds
and requires at most 22 iterations for the web servers.

To confirm that our noise model works as intended, we additionally analyzed the results regarding
the noise they contain.
X-noise is more critical than R-noise as it removes the target page from the candidate list of the iteration for which it happens.
\autoref{tab:proc_accesses} shows the probability that X-noise occurs in an iteration for all noise levels.
The table additionally shows the average size of $X_i$ and $R_i$ recordings.
The results show that our noise model introduces significant X-noise
in up to $90\%$ of the $X_i$ recordings and strongly increases the size of
recordings.

\begin{table}[tb]\centering
\caption{Number of iterations and time required until top set converges ($|T| \leq 5$) for different noise levels.}
\label{tab:time_finished}
\begin{tabular}{cccc}\toprule
	\textbf{Noise Level} & \textbf{Apache} & \textbf{Nginx} & \textbf{OpenSSH}\\ \midrule
       20 & 10 (7.4 s) & 8 (5.56 s) & 21 (38.85 s) \\
       30 & 10 (7.5 s) & 9 (6.62 s) & 42 (85.47 s) \\
       40 & 12 (9.7 s) & 13 (13.2 s) & 46 (111.09 s) \\
       50 & 22 (23.0 s) & 16 (17.84 s) & >100 (>5 min) \\
\bottomrule
\end{tabular}
\end{table}

\begin{table}[tb]\centering
\caption{X-noise probability and average recording size for different noise levels.}
\label{tab:proc_accesses}
\begin{tabular}{cccc}\toprule
	\textbf{Noise Level} & \textbf{Apache} & \textbf{Nginx} & \textbf{OpenSSH}\\ \midrule
       20 & 35\% (8,220) & 34\% (8,355) & 63\% (18,960) \\
       30 & 49\% (10,860) & 51\% (10,360) & 78\% (21,475) \\
       40 & 60\% (13,040) & 62\% (12,430) & 85\% (23,015) \\
       50 & 74\% (15,950) & 69\% (15,970) & 90\% (24,990) \\
\bottomrule
\end{tabular}
\end{table}

Summarizing, our identification mechanism's design enables quick resource identification and is robust against noise and
thus applicable in real-world scenarios. 
Note that the resource requests can also be executed from several distributed
clients to further hide the attack.
Among the candidate pages in the top set, we \emph{always} identified the last commonly accessed
page to contain the target resource for all services.
This observation is comprehensible, as all services must perform operations,
such as opening a socket, before transmitting the requested content.
Together with our converged top set, those observations
allowed us in our test cases to \emph{always} correctly identify our target resource's single
page making the attack very reliable to this point.

\subsection{Data Extraction}

With the knowledge about the location of the resource, we were able to
reliably extract the entire memory of the target \gls{VM} on our prototype
implementation as described in \autoref{sec:design:extraction}.
The resource was always sticky (\autoref{sec:design:resource_stickiness}) over the whole process.
While preserving the \gls{VM}'s stability at all times, the
extraction of its entire 2 GB also worked under the noise model introduced for
the identification phase.
\autoref{tab:extraction_perf} summarizes the extraction speed with different
services for our single-page resource. With OpenSSH, we experienced a higher response time reducing the
extraction speed for this case.
A single-paged resource represents a worst-case scenario, which can be
significantly improved in practice when identifying a larger resource.
This requires fewer
requests to the target \gls{VM} while receiving larger chunks of main memory.

\begin{table}[tb]\centering
\caption{Extraction speed for different target services using a one page-sized
	resource.}
\label{tab:extraction_perf}
\begin{tabular}{ccc}\toprule
	\textbf{Apache} & \textbf{Nginx} & \textbf{OpenSSH}\\ \midrule
       79.4 \nicefrac{KB}{sec} & 79.4 \nicefrac{KB}{sec} & 41.6 \nicefrac{KB}{sec}\\
\bottomrule
\end{tabular}
\end{table}

\subsection{Discussion} 
\label{sec:discussion}

Our evaluation shows that \name is feasible in practice and that it can be used to extract
the entire memory from a SEV-protected \gls{VM} within reasonable time. %subverting the protection of AMD \gls{SEV}.
The results specifically show that critical aspects, such as noise during the identification and the resource
stickiness are managed well by \name.

Nevertheless, there are multiple possibilities for future work to further improve
\name.
For example, in various situations, a full memory dump is not required.
An attacker could consider only extracting the private key of a
web server. This key is likely to be found among the pages accessed
during a request, as it would have to be accessed by the web server
process in order to initiate an encrypted connection. With this knowledge, an
attacker could limit the amount of pages possibly containing the TLS key to a fraction of the
\gls{VM}'s memory, drastically reducing the extraction time.
A similar approach could be performed, for example, with
password hashes or disk encryption keys a service accesses in the course of a request.
An optimization to \name could be made by continuously analyzing the received data during
the extraction phase. As soon as the targeted secret is found,
the attacker could stop the extraction, decreasing the duration of
the attack, increasing its stealthiness.

\section{Countermeasures}
\label{sec:countermeasures}

\name depends on both the possibility to
track the \gls{VM}'s accesses to \glspl{GPA} and the missing integrity protection.
To prevent page faults from leaking information, Shinde et al.
\cite{shinde2016preventing} proposed a method where processes create a
deterministic sequence of page faults, independent from the input. 
However, this is not sufficient to hide the \gls{VM}'s accesses to the critical
resource from the \gls{HV}.
The best a \gls{VM} can achieve is to generate additional page faults to complicate
the resource identification phase.
Additionally, integrity protection can hardly be achieved in software as
the \gls{VM} would require efficient and reliable software mechanisms to protect itself
from modification of memory mappings and contents, e.g., by maintaining hashes in a safe location.
Both mechanisms seem hard to realize to reliably protect an entire \gls{VM} at all times,
and would probably incur an intolerable performance overhead.
We thus consider software-based countermeasures insufficient solutions against our attack.
Therefore, a modification of AMD \gls{SEV} seems inevitable to fully prevent
\name. The best solution seems to be to provide a full-featured integrity and freshness protection
of guest-pages additional to the encryption, as realized in Intel SGX.
However, this likely comes with a high silicon cost to protect full \glspl{VM} compared to SGX enclaves.
A low-cost efficient solution could be to securely combine the hash of the page's content
with the guest-assigned \gls{GPA}. This ensures that pages can not easily be swapped
by changing the \gls{GPA} to \gls{HPA} mapping. Adding a nonce additionally ensures
that an old page for the \gls{GPA} cannot be replayed into the guest by a malicious \gls{HV}.
Integration of such an approach into AMD \gls{SEV} could effectively prevent remapping.

Note that the not yet available extension \gls{SEV-ES} does \emph{not} protect against
\name, since our attack does not require access to any \gls{VM} state encrypted by
\gls{SEV-ES}.

\section{Related Work}
\label{sec:related_work}

While Payer \cite{payer16amd} already pointed out the general problem of the
\gls{SEV} approach when it was announced, 
to the best of our knowledge, only one paper presenting an actual attack on \gls{SEV}
has been published at the time of writing. 
Hetzelt and Buhren \cite{hetzelt2017security} changed the program flow of an SSH service running in an \gls{SEV}-encrypted \gls{VM}.
Using page remapping, they were able to gain unauthorized access to the service.
This requires manual analysis of the SSH service to recognize page access patterns for different data flows.
Our approach does neither require thorough analysis of a specific target service, nor to alter the program flow within the target \gls{VM}.
This eases the application of our method for different services and sustains the target \gls{VM}'s code integrity.
Also, their method requires data being located at certain offsets within a page reducing the probability of a successful attack.
In contrast, our results point out the particularly high success probability of \name even in realistic scenarios where the \gls{VM} is under high load.
Further, they also require a victim logging into his user account in order to be able to remap his session information to the attacker's session who logs in shortly after.
In comparison, \name does not depend on any interaction from a victim, as we can perform all requests necessary for our attack ourselves remotely and at any time.

Buhren et al. \cite{buhren2017fault} leveraged fault attacks on \gls{SME} platforms, which \gls{SEV} is built upon.
They showed that it is possible to extract the private RSA key of a GnuPG user from encrypted memory.
Their attacker model requires the attacker to have control over a process running on the target system and to have physical access to the system.
They first used the process on the target system to perform cache timing attacks in order to identify relevant assets of a GnuPG process in memory.
In the next step, they made use of \gls{DMA} to inject faults into those assets, which caused GnuPG to create faulty signatures.
They used these faulty signatures to calculate parts of an RSA key offline.
Their very specific concept solely applies to \gls{SME} platforms, but not to \glspl{VM} on \gls{SEV} environments and is difficult to execute on productive environments.
Our approach makes it possible to extract the whole memory of a \gls{VM} without physical access and without particular knowledge or control over processes on the target \gls{VM}.

Additionally, in contrast to our prototype, both \cite{hetzelt2017security} and \cite{buhren2017fault} did not realize the attack on real hardware, as only the AMD specifications for \gls{SEV} and \gls{SME} were public.
The results of our work strongly indicate that \cite{hetzelt2017security} also works on real \gls{SEV} platforms.

\section{Conclusion}
\label{sec:concl}

We presented the design and implementation of \name, an attack that reliably
extracts the full plaintext memory of \glspl{VM} encrypted with AMD \gls{SEV}
from a malicious \gls{HV}.
The only major requirement for our method is the presence of a service in the
\gls{VM}, which provides a resource to the outside.
Such services are usually easy to find, since \glspl{VM} are typically and
widely used in server contexts where they host web servers and other remotely
accessible services.
We demonstrated the feasibility of our approach by realizing a prototype on a recent AMD \gls{SEV}-enabled platform.
We evaluated the prototype with different services, namely the Apache and
nginx web servers and an OpenSSH server.
For every service, we considered various levels of concurrent accesses to evaluate \name
under different, realistic load conditions.
In all cases, we were able to efficiently identify the relevant resource of
the target \gls{VM} in memory by analyzing the \gls{VM}'s memory access
patterns from the \gls{HV}.
With the gained knowledge we were able to use the malicious \gls{HV} to remap the
resource to other memory pages and to iteratively request all the \gls{VM}'s
memory in reasonable time.
As \name is independent of the specific service, our method can easily be
adapted to a variety of different attack scenarios in practice.
\name demonstrates that a malicious \gls{HV} still remains able to extract
sensitive data from its SEV-enabled guest \glspl{VM}.

\section*{Acknowledgements}

This work has been partially funded in the project CAR-BITS.de by the German Federal Ministry for Economic Affairs and Energy under the reference 01MD16004B.

%% file: paper-eurosec.bbl

\begin{thebibliography}{15}

%%% the Web address.

\ifx \showCODEN    \undefined \def \showCODEN     #1{\unskip}     \fi
\ifx \showDOI      \undefined \def \showDOI       #1{#1}\fi
\ifx \showISBNx    \undefined \def \showISBNx     #1{\unskip}     \fi
\ifx \showISBNxiii \undefined \def \showISBNxiii  #1{\unskip}     \fi
\ifx \showISSN     \undefined \def \showISSN      #1{\unskip}     \fi
\ifx \showLCCN     \undefined \def \showLCCN      #1{\unskip}     \fi
\ifx \shownote     \undefined \def \shownote      #1{#1}          \fi
\ifx \showarticletitle \undefined \def \showarticletitle #1{#1}   \fi
\ifx \showURL      \undefined \def \showURL       {\relax}        \fi
% invisible to TeX
\providecommand\bibfield[2]{#2}
\providecommand\bibinfo[2]{#2}
\providecommand\natexlab[1]{#1}
\providecommand\showeprint[2][]{arXiv:#2}

\bibitem[\protect\citeauthoryear{{Advanced Micro Devices}}{{Advanced Micro
  Devices}}{2017a}]%
        {AMDSEV}
\bibfield{author}{\bibinfo{person}{{Advanced Micro Devices}}.}
  \bibinfo{year}{2017}\natexlab{a}.
\newblock \bibinfo{title}{{GitHub - AMDESE/AMDSEV: AMD Secure Encrypted
  Virtualization}}.
\newblock \bibinfo{howpublished}{\url{https://github.com/AMDESE/AMDSEV}}.
\newblock
\newblock
\shownote{[Online; accessed 2018-03-04].}


\bibitem[\protect\citeauthoryear{{Advanced Micro Devices}}{{Advanced Micro
  Devices}}{2017b}]%
        {AMD2017API}
\bibfield{author}{\bibinfo{person}{{Advanced Micro Devices}}.}
  \bibinfo{year}{2017}\natexlab{b}.
\newblock \bibinfo{title}{{Secure Encrypted Virtualization API Version 0.14}}.
\newblock
  \bibinfo{howpublished}{\url{http://support.amd.com/TechDocs/55766_SEV-KM\%20API_Specification.pdf}}.
\newblock
\newblock
\shownote{[Online; accessed 2018-03-04].}


\bibitem[\protect\citeauthoryear{Becher, Dornseif, and Klein}{Becher
  et~al\mbox{.}}{2005}]%
        {becher2005firewire}
\bibfield{author}{\bibinfo{person}{Michael Becher},
  \bibinfo{person}{Maximillian Dornseif}, {and} \bibinfo{person}{Christian~N
  Klein}.} \bibinfo{year}{2005}\natexlab{}.
\newblock \showarticletitle{{FireWire: All Your Memory Are Belong To Us}}.
\newblock \bibinfo{journal}{\emph{Proceedings of CanSecWest}}
  (\bibinfo{year}{2005}).
\newblock


\bibitem[\protect\citeauthoryear{Boileau}{Boileau}{2006}]%
        {boileau2006hit}
\bibfield{author}{\bibinfo{person}{Adam Boileau}.}
  \bibinfo{year}{2006}\natexlab{}.
\newblock \showarticletitle{{Hit by a bus: Physical access attacks with
  Firewire}}.
\newblock \bibinfo{journal}{\emph{Presentation, Ruxcon}}
  (\bibinfo{year}{2006}).
\newblock


\bibitem[\protect\citeauthoryear{Buhren, Gueron, Nordholz, Seifert, and
  Vetter}{Buhren et~al\mbox{.}}{2017}]%
        {buhren2017fault}
\bibfield{author}{\bibinfo{person}{Robert Buhren}, \bibinfo{person}{Shay
  Gueron}, \bibinfo{person}{Jan Nordholz}, \bibinfo{person}{Jean-Pierre
  Seifert}, {and} \bibinfo{person}{Julian Vetter}.}
  \bibinfo{year}{2017}\natexlab{}.
\newblock \showarticletitle{{Fault Attacks on Encrypted General Purpose Compute
  Platforms}}. In \bibinfo{booktitle}{\emph{Proceedings of the Seventh ACM on
  Conference on Data and Application Security and Privacy}}. ACM,
  \bibinfo{pages}{197--204}.
\newblock


\bibitem[\protect\citeauthoryear{Devine and Vissian}{Devine and
  Vissian}{2009}]%
        {pcie}
\bibfield{author}{\bibinfo{person}{Christophe Devine} {and}
  \bibinfo{person}{Guillaume Vissian}.} \bibinfo{year}{2009}\natexlab{}.
\newblock \showarticletitle{{Compromission physique par le bus PCI}}.
\newblock \bibinfo{journal}{\emph{Proceedings of SSTIC}}
  (\bibinfo{year}{2009}).
\newblock


\bibitem[\protect\citeauthoryear{Guangrong}{Guangrong}{2016}]%
        {guangrong2016patch}
\bibfield{author}{\bibinfo{person}{Xiao Guangrong}.}
  \bibinfo{year}{2016}\natexlab{}.
\newblock \bibinfo{title}{{[PATCH v3 00/11] KVM: x86: track guest page
  access}}.
\newblock
  \bibinfo{howpublished}{\url{http://www.mail-archive.com/linux-kernel@vger.kernel.org/msg1076006.html}}.
\newblock
\newblock
\shownote{[Online; accessed 2018-03-04].}


\bibitem[\protect\citeauthoryear{Halderman, Schoen, Heninger, Clarkson, Paul,
  Calandrino, Feldman, Appelbaum, and Felten}{Halderman et~al\mbox{.}}{2009}]%
        {Halderman08lestwe}
\bibfield{author}{\bibinfo{person}{J~Alex Halderman}, \bibinfo{person}{Seth~D
  Schoen}, \bibinfo{person}{Nadia Heninger}, \bibinfo{person}{William
  Clarkson}, \bibinfo{person}{William Paul}, \bibinfo{person}{Joseph~A
  Calandrino}, \bibinfo{person}{Ariel~J Feldman}, \bibinfo{person}{Jacob
  Appelbaum}, {and} \bibinfo{person}{Edward~W Felten}.}
  \bibinfo{year}{2009}\natexlab{}.
\newblock \showarticletitle{{Lest We Remember: Cold-boot Attacks on Encryption
  Keys}}.
\newblock \bibinfo{journal}{\emph{Commun. ACM}} (\bibinfo{year}{2009}),
  \bibinfo{pages}{91--98}.
\newblock


\bibitem[\protect\citeauthoryear{Hetzelt and Buhren}{Hetzelt and
  Buhren}{2017}]%
        {hetzelt2017security}
\bibfield{author}{\bibinfo{person}{Felicitas Hetzelt} {and}
  \bibinfo{person}{Robert Buhren}.} \bibinfo{year}{2017}\natexlab{}.
\newblock \showarticletitle{{Security Analysis of Encrypted Virtual Machines}}.
  In \bibinfo{booktitle}{\emph{Proceedings of the 13th ACM SIGPLAN/SIGOPS
  International Conference on Virtual Execution Environments}}. ACM,
  \bibinfo{pages}{129--142}.
\newblock


\bibitem[\protect\citeauthoryear{Kaplan, Powell, and Woller}{Kaplan
  et~al\mbox{.}}{2016}]%
        {kaplan2016amd}
\bibfield{author}{\bibinfo{person}{David Kaplan}, \bibinfo{person}{Jeremy
  Powell}, {and} \bibinfo{person}{Tom Woller}.}
  \bibinfo{year}{2016}\natexlab{}.
\newblock \bibinfo{booktitle}{\emph{{AMD memory encryption}}}.
\newblock \bibinfo{type}{{T}echnical {R}eport}. \bibinfo{institution}{Advanced
  Micro Devices}.
\newblock


\bibitem[\protect\citeauthoryear{{Microsoft}}{{Microsoft}}{2017}]%
        {Microsoft2017Vuln}
\bibfield{author}{\bibinfo{person}{{Microsoft}}.}
  \bibinfo{year}{2017}\natexlab{}.
\newblock \bibinfo{title}{{Microsoft Security Bulletin MS17-008 - Critical}}.
\newblock
  \bibinfo{howpublished}{\url{https://technet.microsoft.com/en-us/library/security/ms17-008.aspx}}.
\newblock
\newblock
\shownote{[Online; accessed 2018-03-04].}


\bibitem[\protect\citeauthoryear{Payer}{Payer}{2016}]%
        {payer16amd}
\bibfield{author}{\bibinfo{person}{Mathias Payer}.}
  \bibinfo{year}{2016}\natexlab{}.
\newblock \bibinfo{title}{{AMD SEV attack surface: a tale of too much trust}}.
\newblock
  \bibinfo{howpublished}{\url{https://nebelwelt.net/blog/20160922-AMD-SEV-attack-surface.html}}.
\newblock
\newblock
\shownote{[Online; accessed 2018-03-04].}


\bibitem[\protect\citeauthoryear{Shinde, Chua, Narayanan, and Saxena}{Shinde
  et~al\mbox{.}}{2016}]%
        {shinde2016preventing}
\bibfield{author}{\bibinfo{person}{Shweta Shinde}, \bibinfo{person}{Zheng~Leong
  Chua}, \bibinfo{person}{Viswesh Narayanan}, {and} \bibinfo{person}{Prateek
  Saxena}.} \bibinfo{year}{2016}\natexlab{}.
\newblock \showarticletitle{{Preventing page faults from telling your
  secrets}}. In \bibinfo{booktitle}{\emph{Proceedings of the 11th ACM on Asia
  Conference on Computer and Communications Security}}. ACM,
  \bibinfo{pages}{317--328}.
\newblock


\bibitem[\protect\citeauthoryear{{VMware}}{{VMware}}{2017}]%
        {VMWare2017Vuln}
\bibfield{author}{\bibinfo{person}{{VMware}}.} \bibinfo{year}{2017}\natexlab{}.
\newblock \bibinfo{title}{{VMSA-2017-0006: VMware ESXi, Workstation and Fusion
  updates address critical and moderate security issues}}.
\newblock
  \bibinfo{howpublished}{\url{https://www.vmware.com/security/advisories/VMSA-2017-0006.html}}.
\newblock
\newblock
\shownote{[Online; accessed 2018-03-04].}


\bibitem[\protect\citeauthoryear{{Xenproject.org Security
  Team}}{{Xenproject.org Security Team}}{2017}]%
        {Xen2017Vuln}
\bibfield{author}{\bibinfo{person}{{Xenproject.org Security Team}}.}
  \bibinfo{year}{2017}\natexlab{}.
\newblock \bibinfo{title}{{x86: broken check in memory\_exchange() permits PV
  guest breakout}}.
\newblock
  \bibinfo{howpublished}{\url{https://xenbits.xen.org/xsa/advisory-212.html}}.
\newblock
\newblock
\shownote{[Online; accessed 2018-03-04].}


\end{thebibliography}
